\documentclass[twocolumn,twocolappendix]{aastex63}

\newcommand\alfven{Alfv\'{e}n}

\usepackage{amsmath}
\usepackage{mathrsfs}
\allowdisplaybreaks

\submitjournal{ApJ}

\shorttitle{Alfv\'{e}n wave nonlinear breakout}
\shortauthors{Yuan et al.}
\graphicspath{{./}{figures/}}

\begin{document}

\title{Magnetar bursts due to Alfv\'{e}n wave nonlinear breakout}

\correspondingauthor{Yajie Yuan}
\email{yyuan@flatironinstitute.org}

\author[0000-0002-0108-4774]{Yajie Yuan}
\affiliation{Center for Computational Astrophysics, Flatiron Institute, 162 Fifth Avenue, New York, NY 10010, USA}

\author[0000-0001-5660-3175]{
Andrei M. Beloborodov
}
\affil{
Physics Department and Columbia Astrophysics Laboratory, Columbia University, 538  West 120th Street New York, NY 10027}
\affil{
Max Planck Institute for Astrophysics, Karl-Schwarzschild-Str. 1, D-85741, Garching, Germany
}

\author[0000-0002-4738-1168]{Alexander Y. Chen}
\affil{JILA, University of Colorado, 440 UCB, Boulder, CO 80309, USA}

\author{Yuri Levin}
\affiliation{Center for Computational Astrophysics, Flatiron Institute, 162 Fifth Avenue, New York, NY 10010, USA}
\affil{
Physics Department and Columbia Astrophysics Laboratory, Columbia University, 538  West 120th Street New York, NY 10027}
\affil{Department of Physics and Astronomy, Monash University, Clayton VIC 3800, Australia}

\author[0000-0002-0491-1210]{Elias R. Most}
\affil{Princeton Center for Theoretical Science, Princeton University, Princeton, NJ 08544, USA}
\affil{Princeton Gravity Initiative, Princeton University, Princeton, NJ 08544, USA}
\affil{School of Natural Sciences, Institute for Advanced Study, Princeton, NJ 08540, USA}

\author[0000-0001-7801-0362]{Alexander A. Philippov}
\affiliation{Center for Computational Astrophysics, Flatiron Institute, 162 Fifth Avenue, New York, NY 10010, USA}
\affil{Department of Physics, University of Maryland, College Park, MD 20742, USA}

\begin{abstract}
The most common form of magnetar activity is short X-ray bursts, with durations from milliseconds to seconds, and luminosities ranging from $10^{36}$ to $10^{43}\ {\rm erg}\,{\rm s}^{-1}$. Recently, an X-ray burst from the galactic magnetar SGR 1935+2154 was detected to be coincident with two fast radio burst (FRB) like events from the same source, providing evidence that FRBs may be linked to magnetar bursts. Using fully 3D force-free electrodynamics simulations, we show that such magnetar bursts may be produced by \alfven{} waves launched from localized magnetar quakes: a wave packet propagates to the outer magnetosphere, becomes nonlinear, and escapes the magnetosphere, forming an ultra-relativistic ejecta. 
The ejecta pushes open the magnetospheric field lines, creating current sheets behind it. Magnetic reconnection can happen at these current sheets, leading to plasma energization and X-ray emission. The angular size of the ejecta can be compact, $\lesssim 1$ sr if the quake launching region is small, $\lesssim 0.01$ sr at the stellar surface.  We discuss implications for the FRBs and the coincident X-ray burst from SGR 1935+2154.

\end{abstract}



\section{Introduction} \label{sec:intro}
Magnetars are young, strongly magnetized neutron stars with surface magnetic fields reaching $B\sim 10^{14}\,\mathrm{G}$, beyond the quantum critical field $B_c=4.4\times10^{13}$ G (\cite{1992ApJ...392L...9D}, see \cite{2017ARA&A..55..261K} for a recent review). Their quiescent X-ray luminosity is usually far larger than the spin down luminosity, so the emission is believed to be powered by dissipation of the strong magnetic field instead of rotation. They often display dramatic variability in the X-ray and soft $\gamma$-ray band.
This activity includes short (milliseconds to seconds duration) ``bursts'' with peak X-ray luminosity ranging from $10^{36}$ to $10^{43}\ {\rm erg\, s}^{-1}$, much longer (weeks to months) ``outbursts'', and sometimes ``giant flares'' with sudden release of $> 10^{44}$ erg of energy.

The short bursts are by far the most common type of magnetar activities. \citet{1995MNRAS.275..255T} first proposed a picture for magnetar bursts: the internal field evolution could build up stress locally on the neutron star crust; the stress could become strong enough to cause mechanical failure of the crust, which leads to a sudden shift in the magnetospheric footpoints; this sends \alfven{} waves into the magnetosphere, and the subsequent dissipation of the \alfven{} waves in the magnetosphere could power the X-ray emission. However, the dissipation mechanism and the accompanying radiative processes are not established. The radius of burst emission is also unknown.

Magnetars have also been proposed as sources of mysterious fast radio bursts (FRBs)---the bright, millisecond-long GHz bursts detected from cosmological distances \citep[e.g.,][]{2013Sci...341...53T}. Recent detection of FRB-like bursts from a galactic magnetar, SGR 1935+2154, provides evidence that magnetars are indeed capable of producing at least some of the FRBs \citep{2020Natur.587...54T,2020Natur.587...59B}. The radio bursts were detected during an active period of the magnetar, and were coincident with an X-ray burst of energy $\sim 10^{40}$ erg \citep{2020ApJ...898L..29M,2021NatAs...5..372R,2021NatAs...5..378L}. This detection provides a good opportunity to understand more about magnetar activity and its relation to FRBs.

Our previous work \citep{2020ApJ...900L..21Y} proposed a possible scenario for the simultaneous generation of the X-ray and radio bursts from SGR 1935+2154. Using 2D axisymmetric simulations, we showed that low-amplitude \alfven{} waves from a magnetar quake may propagate to the outer magnetosphere, become nonlinear and convert to ``plasmoids'' (closed magnetic loops) that accelerate away from the star. An \alfven{} wave packet with an energy $\mathscr{E}_A\sim10^{40}$ erg, as required by the energetics of the X-ray burst from SGR 1935+2154, forms freely expanding ejecta at a radius $R\sim10^8$\,cm, where the wave energy exceeds the local magnetospheric energy. The ejecta pushes out the magnetospheric field lines, and a current sheet forms behind it, leading to magnetic reconnection. Such reconnection events must produce X-ray emission. The spectrum of the resulting X-ray burst was calculated by Beloborodov (2021) and showed good agreement with observations.

Magnetospheric ejecta play a significant role in FRB models. They were proposed to
launch blast waves in the magnetar wind \citep{2017ApJ...843L..26B}, which are capable of emitting coherent radio waves by the synchrotron maser process \citep[e.g.][]{2014MNRAS.442L...9L,2017ApJ...843L..26B,2020ApJ...896..142B,2019MNRAS.485.3816P,2021PhRvL.127c5101S}.
GHz waves can also be seeded by the process of magnetic reconnection triggered by magnetospheric ejecta \citep{2019MNRAS.483.1731L,2019ApJ...876L...6P,2020ApJ...900L..21Y,2020ApJ...897....1L,2022arXiv220304320M}; these waves may be released by the ejecta when it expands to a large radius.

In the present paper, we extend the axisymmetric simulations of \cite{2020ApJ...900L..21Y} to a full 3D model of a magnetospheric explosion from a magnetar quake. We still use the framework of force-free electrodynamics (FFE) designed for magnetically-dominated systems, such as magnetospheres of neutron stars. FFE is essentially the infinite magnetization limit of magnetohydrodynamics; the plasma is treated as a massless conducting fluid, moving under the electromagnetic stress, while providing the necessary charge and current densities. 
We describe the problem setup and our numerical method in \S\ref{sec:method}, and present the results in \S\ref{sec:result}. 
The results are discussed and compared with the previous 2D simulations in \S\ref{sec:discussion}. 
Our main conclusions are summarized
in \S\ref{sec:conclusion}.

\section{Problem setup and numerical method}\label{sec:method}

We consider a localized star quake that exerts a twisting \alfven{ic} perturbation on a small patch offset from the magnetic pole on the neutron star surface. The neutron star is assumed to have a simple dipole magnetic field. We consider the parameter regime as suitable for SGR 1935+2154. The light cylinder of the magnetar is located at $r_{\rm LC}=c P/(2\pi)\sim 1.6\times 10^{10}$ cm, where $P\approx3.25$ s is the spin period of the magnetar, and we expect the \alfven{} wave to propagate in the closed magnetosphere along a flux tube that extends to a radius $R\sim 100r_*\sim 10^8$ cm \citep{2020ApJ...900L..21Y}, where $r_*$ is the radius of the neutron star. Since $R/r_{\rm LC}\sim10^{-2}$, the rotation induced electric field at $R$ is $E_0\sim (R/r_{\rm LC})B_{0}(R)\ll B_0(R)$, where $B_0$ is the background magnetic field of the magnetar. On the other hand, an \alfven{} wave that becomes nonlinear at $R$ will have a wave electric field $\delta E\gtrsim B_{0}(R)\gg E_0$, therefore, we can neglect the rotation of the neutron star to a good approximation in this study.

Furthermore, for the \alfven{} wave to reach a radius of $R\sim 100 r_*$, the launching region on the stellar surface should be located at a polar angle $\theta\sim0.1$ with respect to the magnetic pole. For our simulation, to cover as much dynamic range as possible, we put our inner boundary at $r_{\rm in}=10r_*$. The evolution of the \alfven{} wave before reaching $r_{\rm in}$ should be purely linear and well described by WKB theory. The wave packet will reach a polar angle $\theta\sim 0.32$ at $r=r_{\rm in}$. Our simulations will then self-consistently track the evolution of the \alfven{} wave beyond $r_{\rm in}$. Our detailed setup is as follows.

\begin{figure}
    \centering
    \includegraphics[width=0.4\columnwidth]{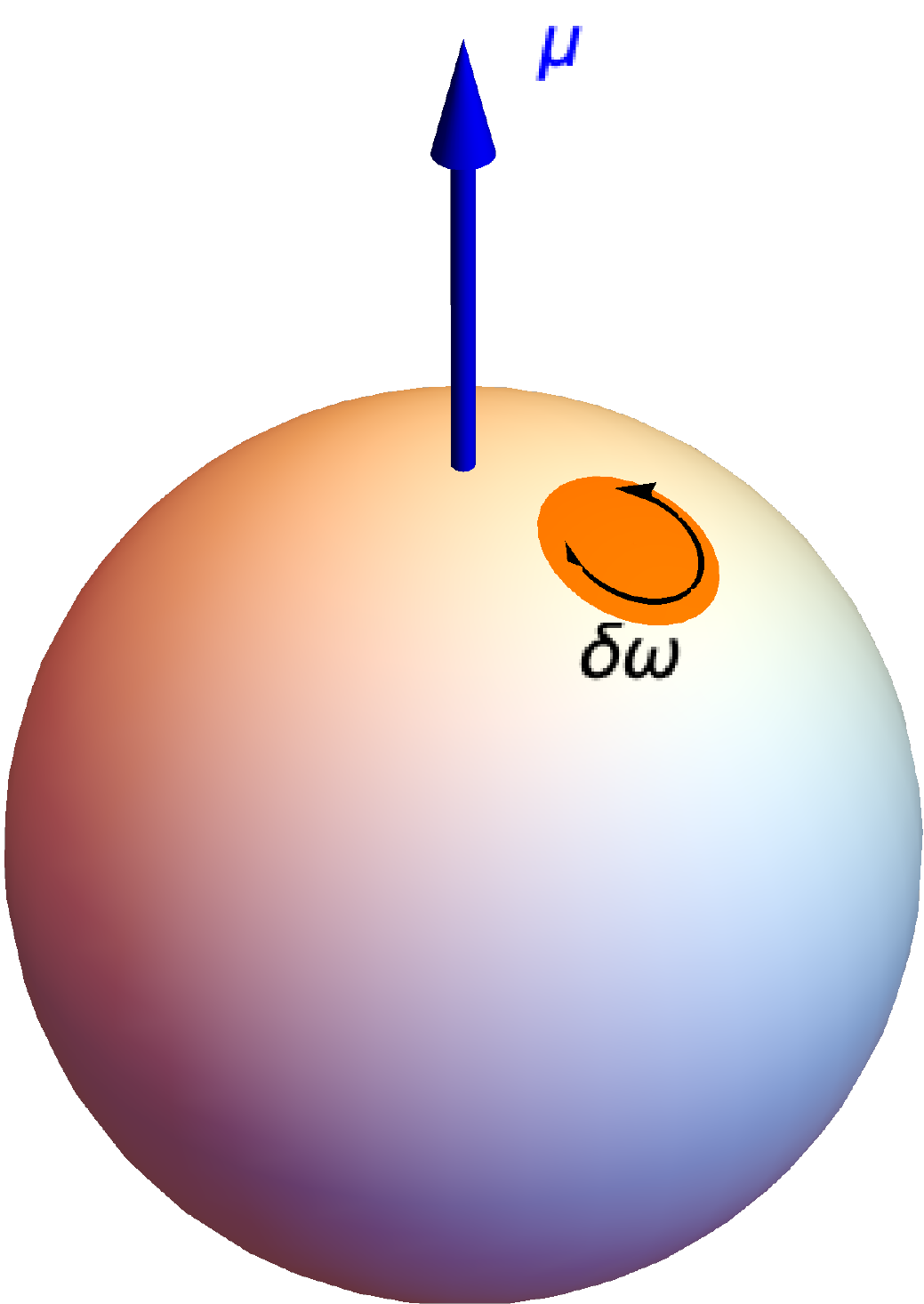}
    \caption{The region on the inner boundary of the simulation domain, $r=r_{\rm in}$, where a twisting perturbation is applied. $\pmb{\mu}$ indicates the magnetic moment of the neutron star.}
    \label{fig:perturbation}
\end{figure}

We introduce the \alfven{} wave perturbation by twisting a small, circular region on the $r=r_{\rm in}$ surface, as shown in  Figure \ref{fig:perturbation}. The circular region is centered at a polar angle $\theta=0.4$ and azimuth angle $\phi=0$, with a radius $r_1=0.2r_{\rm in}$. This motion twists back and forth one foot point of a closed flux bundle, and breaks the axial symmetry of the initial dipole configuration. The twist angular velocity with respect to the twisting center has the following profile
\begin{equation}\label{eq:perturbation}
    \delta\omega=
    \begin{cases}
    \displaystyle
    \delta\omega_0\cos^2\left(\frac{\pi r'}{2 r_1}\right)\sin\left(\frac{2\pi n t}{T}\right)\sin^2\left(\frac{\pi t}{T}\right), & 0\le t\le T,\\
    0, & t>T,
    \end{cases}
\end{equation}
where $\delta\omega_0$ is the amplitude, $r'$ is the distance to the twisting center, $T$ is the duration of the twist, and $n$ determines the number of wave periods. The factor $\cos^2(\pi r'/2 r_1)$ ensures that the perturbation smoothly goes to zero at the boundary of the circular region, while the factor $\sin^2(\pi t/T)$ allows the perturbation to gradually transition to zero at the beginning and the end. We use these smooth profiles to avoid any numerical pathology.

For our simulation domain, we employ a uniform, 3-dimensional Cartesian grid, with the neutron star located at the origin. The inner boundary radius $r_{\rm in}$ is typically resolved by 64 grid points (the highest resolution run uses 128 cells per $r_{\rm in}$ length). At the inner boundary, we enforce the perfectly conducting boundary condition. To avoid the stair stepping at $r_{\rm in}$, we force the fields to known values inside $r_{\rm in}$ with a smoothing kernel \citep{2006ApJ...648L..51S}. The grid covers the region $0\le x\le40$, $-20\le y\le 20$, $-20\le z\le 20$ (lengths are in units of $r_{\rm in}$ and times are in units of $r_{\rm in}/c$, same below). The outer boundaries of the computational grid are covered by an absorbing layer that damps the outgoing waves \citep[e.g.,][]{2015MNRAS.448..606C,2019MNRAS.487.4114Y}.

We use our code \emph{Coffee} \citep[COmputational Force FreE Electrodynamics,][]{2020ApJ...893L..38C}\footnote{\url{https://github.com/fizban007/CoffeeGPU}} to numerically solve the time-dependent force-free equations \citep[e.g.,][]{1999astro.ph..2288G,2002luml.conf..381B}
\begin{align}
  \frac{\partial\mathbf{E}}{\partial t}&= \nabla\times\mathbf{B}-\mathbf{J},\label{eq:FF_dEdt}\\
  \frac{\partial\mathbf{B}}{\partial t}&=- \nabla\times\mathbf{E},\label{eq:FF_dBdt}\\
  \mathbf{J}&=\nabla\cdot\mathbf{E}\frac{\mathbf{E}\times\mathbf{B}}{B^2}+\frac{(\mathbf{B}\cdot\nabla\times\mathbf{B}-\mathbf{E}\cdot\nabla\times\mathbf{E})\mathbf{B}}{B^2},\label{eq:FF_J}
\end{align}
with the constraints $\mathbf{E}\cdot\mathbf{B}=0$ and $E<B$ (we use Heaviside-Lorentz units and set $c=1$). 
A brief summary of the basic algorithms used by {\it Coffee} and the results from convergence tests can be found in Appendix \ref{sec:convergence}.

\begin{figure*}[htb]
    \centering
    \includegraphics[width=\textwidth]{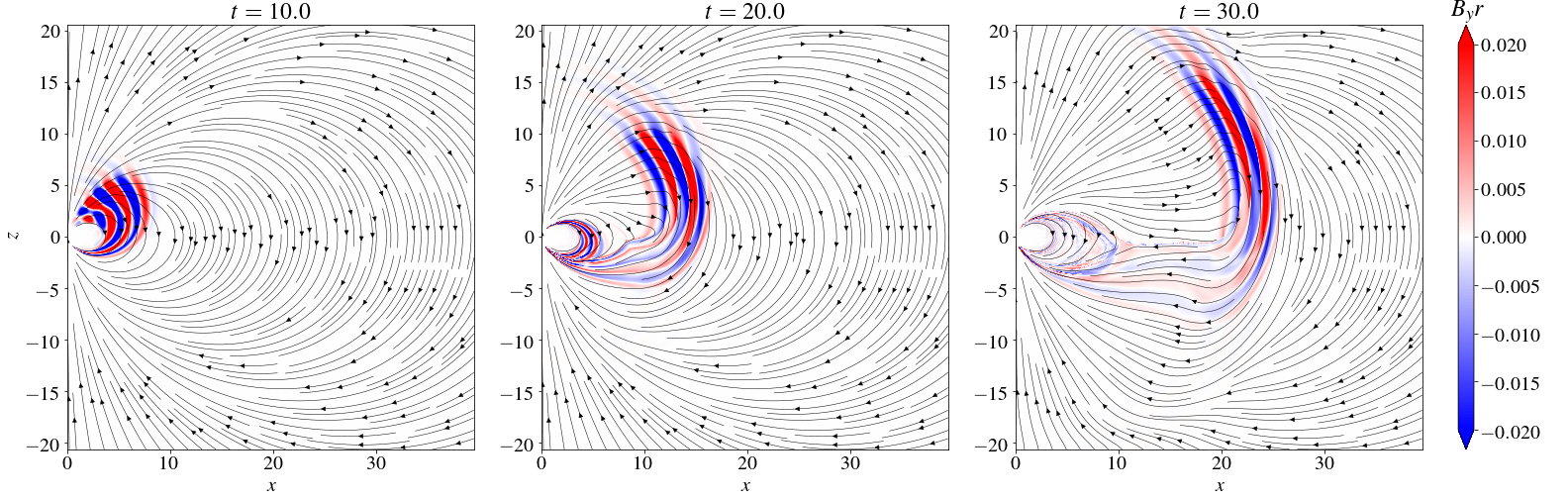}\\
    \includegraphics[width=\textwidth]{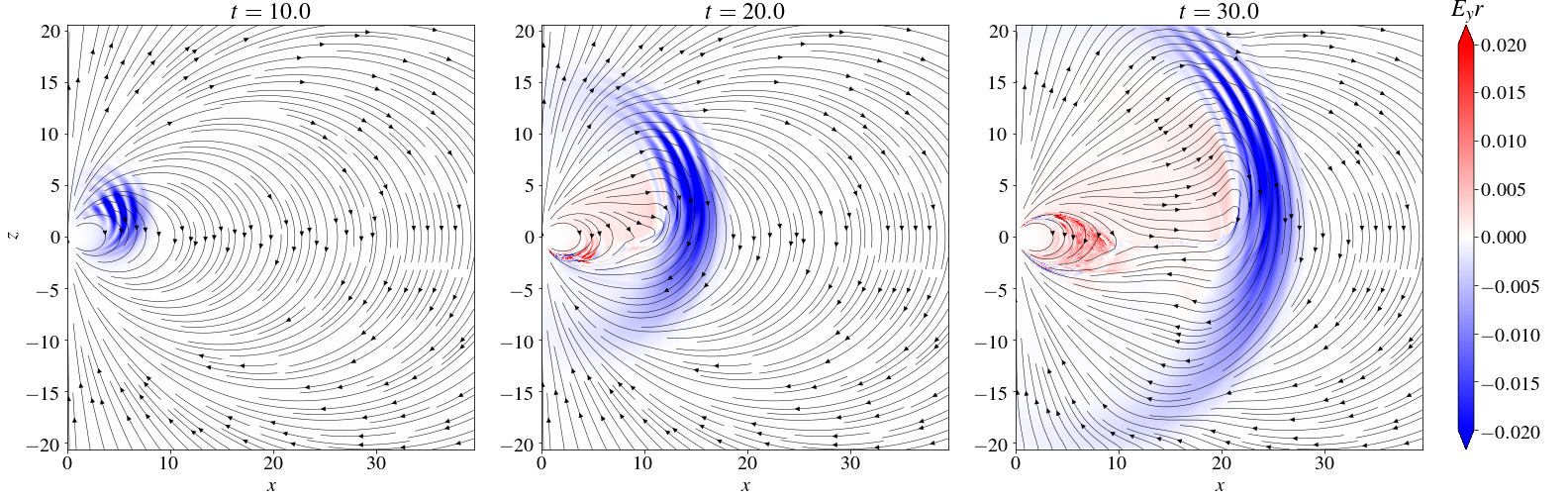}
    \caption{Slices of the electromagnetic field on the $y=0$ ($\phi=0$) plane, at three different time steps. In the top row, color shows the magnetic field component perpendicular to the plane, $B_y$; in the bottom row, color shows the electric field component perpendicular to the plane, $E_y$. In all panels, streamlines show the in-plane magnetic field. Lengths are in units of the inner boundary radius $r_{\rm in}$ and times are in units of $r_{\rm in}/c$ (same below). Note that although we showed $B_y$ and $E_y$ here, the total electric field $\mathbf{E}$ is perpendicular to the total magnetic field $\mathbf{B}$.}
    \label{fig:Bslice}
    \vspace{0.5cm}
    \includegraphics[width=0.32\textwidth]{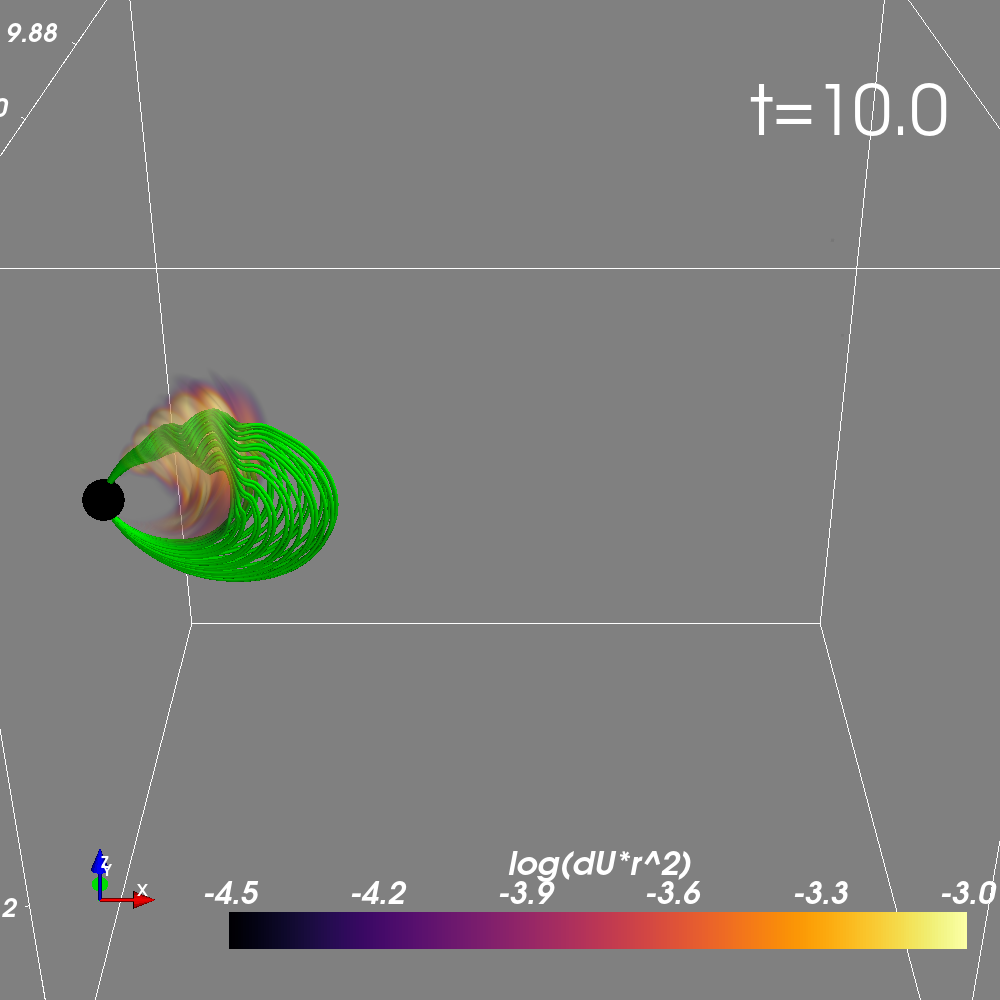}
    \includegraphics[width=0.32\textwidth]{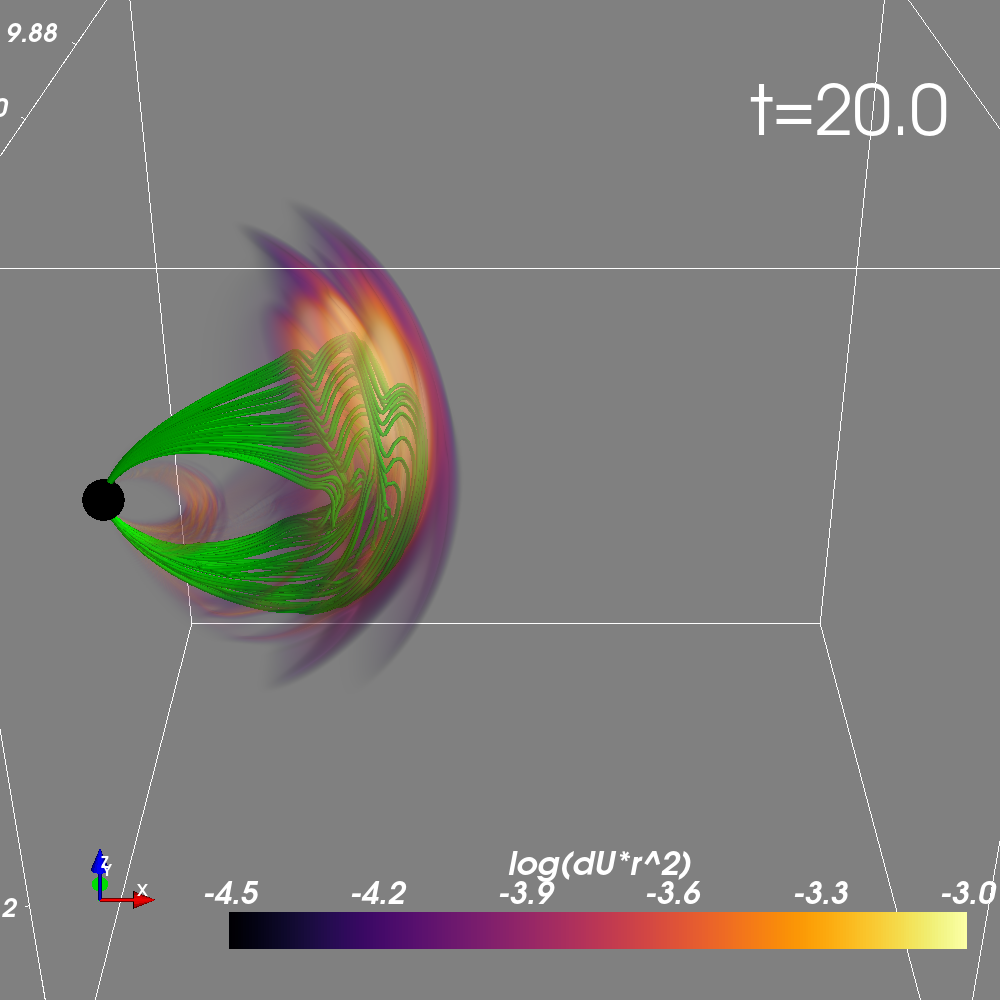}
    \includegraphics[width=0.32\textwidth]{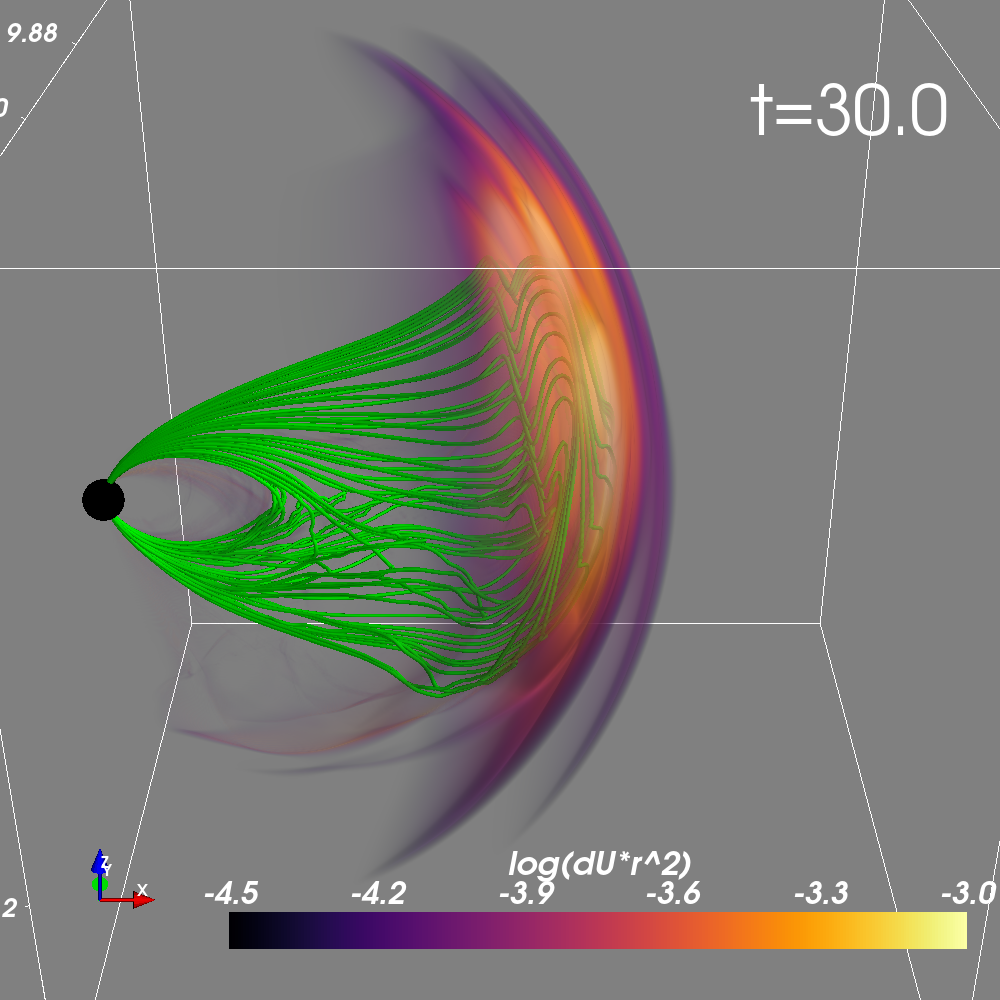}
    \caption{3D rendering of the energy density in the perturbed electromagnetic field $\delta U=(\delta B^2+\delta E^2)/2$, where $\delta \mathbf{B}=\mathbf{B}-\mathbf{B}_0$ is the perturbation magnetic field and $\delta \mathbf{E}$ is the perturbation electric field, at three time steps corresponding to Figure \ref{fig:Bslice}. The green lines are a bundle of field lines within the flux tube perturbed by the \alfven{} wave.}
    \label{fig:dUB_3d}
\end{figure*}

\begin{figure*}[htb]
    \centering
    \includegraphics[width=0.32\textwidth]{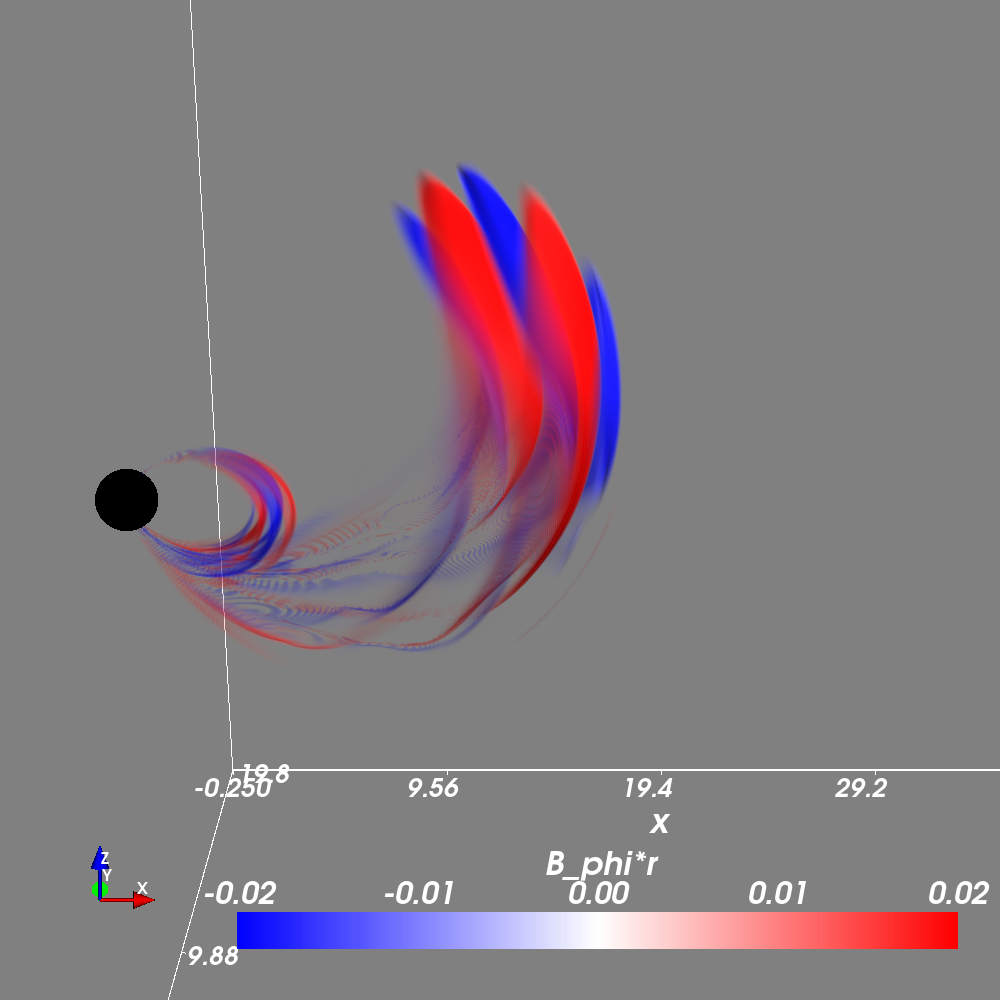}
    \includegraphics[width=0.32\textwidth]{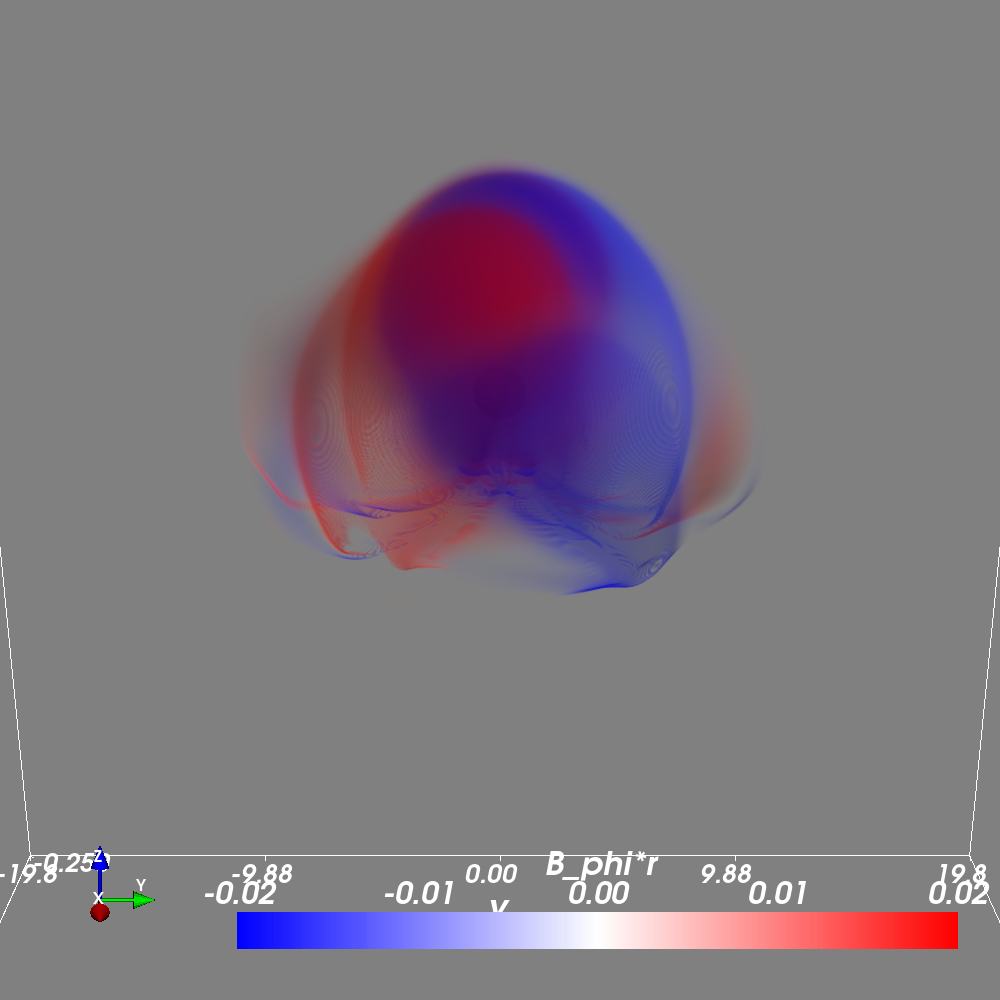}
    \includegraphics[width=0.32\textwidth]{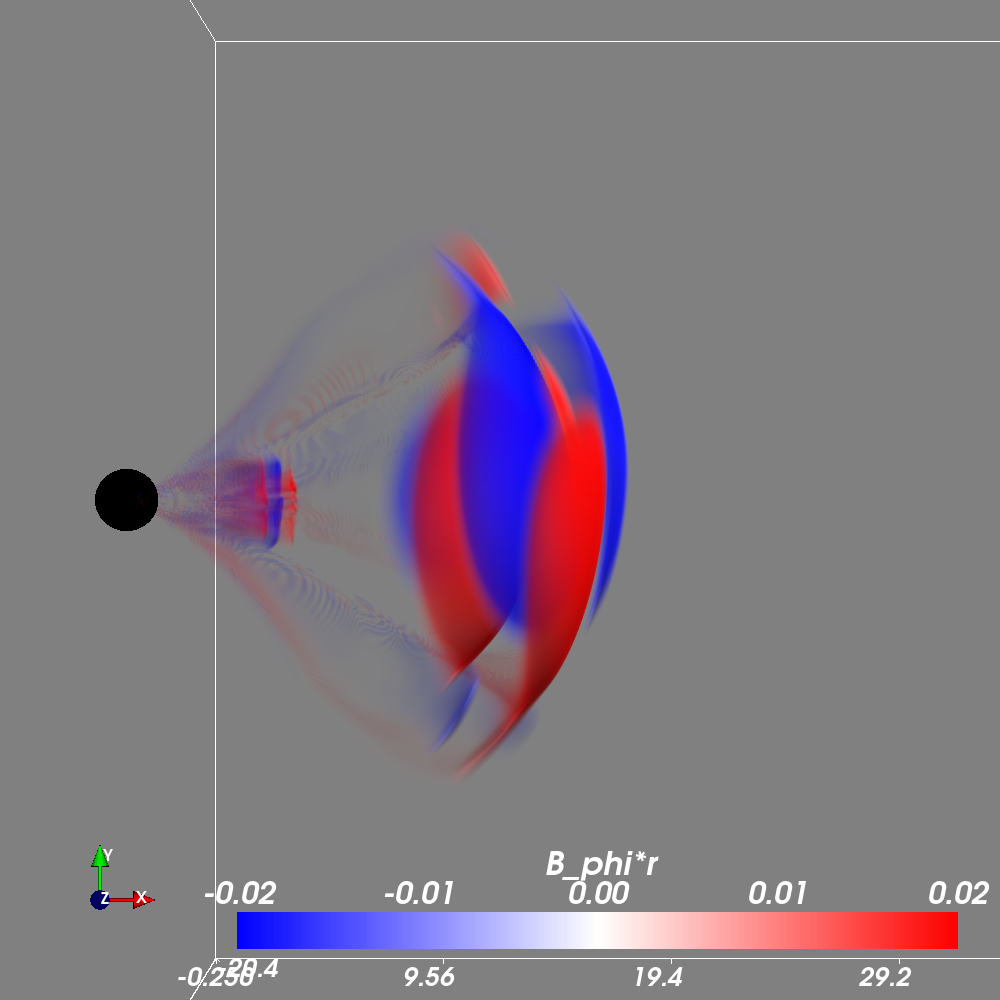}
    \caption{Side view, front view and top view of the toroidal magnetic field $B_{\phi}$ in 3D, at the time step $t=20$.}
    \label{fig:Bphi3d}
\end{figure*}

\begin{figure*}[htb]
    \centering
    \includegraphics[width=0.4\textwidth]{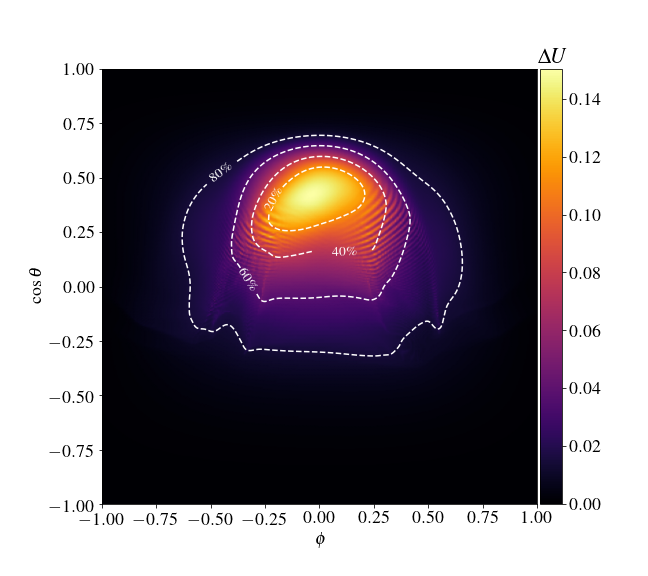}
    \includegraphics[width=0.4\textwidth]{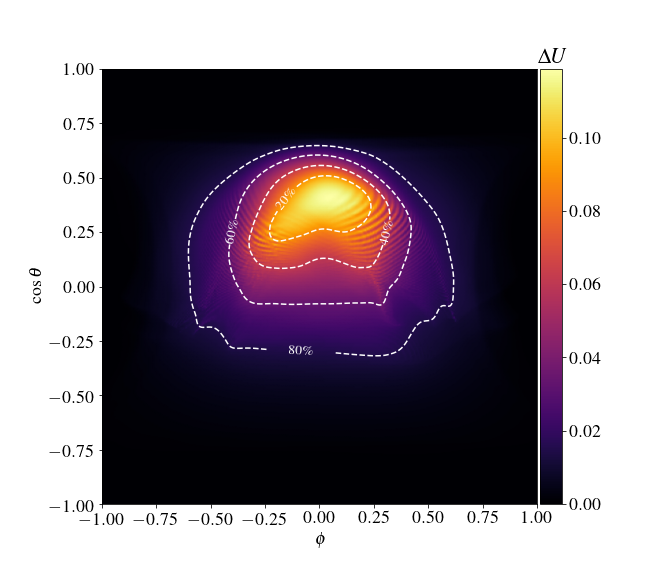}
    \caption{Angular distribution of the electromagnetic energy in the ejecta, measured at $t=25$ (left) and $t=35$ (right). The perturbation electromagnetic energy density has been integrated along the radial direction between two spheres with radius $r=t-T$ and $r=t$ that enclose the ejecta shell. White dashed lines show the 20\%, 40\%, 60\% and 80\% containment regions.}
    \label{fig:dUangular}
\end{figure*}

\begin{figure*}
    \centering
    \includegraphics[width=\textwidth]{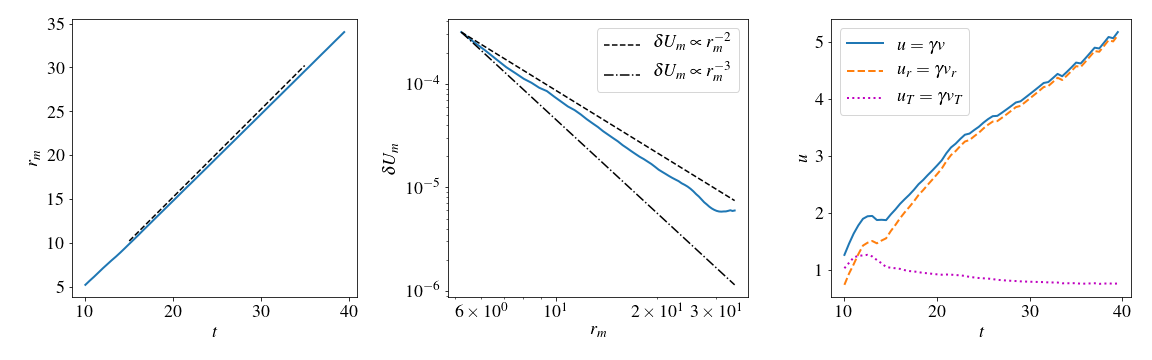}
    \caption{Tracking the evolution of the maximum perturbation electromagnetic energy density in the ejecta, and the drift proper velocity at that point. In the left panel, the blue line shows the distance $r_m$ between the location of maximum $\delta Ur^2$ and the stellar center, as a function of time. The dashed line is a reference of purely radial motion with speed $c$. In the middle panel, the blue line shows $\delta U_m\equiv\delta U(\mathbf{r}_m)$ as a function of $r_m$. The dashed line and dash-dotted line are two different scaling relations for reference. In the right panel, the blue line shows the total magnitude of the drift proper velocity, $u=\gamma v$; the orange dashed line shows its radial component, $u_r=\gamma v_r$; the magenta dotted line shows its transverse component, $u_T=\gamma v_T=\gamma\sqrt{v_{\theta}^2+v_{\phi}^2}$.}
    \label{fig:Um_ud_scaling}
\end{figure*}

\section{Results}\label{sec:result}

Let us first show the results from an example run where the initial \alfven{} wave perturbation has $\delta\omega_0=2.0$, duration $T=10$, and $n=4$ periods. For these parameters, the initial maximum relative amplitude of the wave is $\delta B/B\sim 0.05$ at the inner boundary $r_{\rm in}$, and the total injected energy is $3.4\times10^{-3}\mu^2r_{\rm in}^{-3}/(4\pi)$, where $\mu$ is the magnetic dipole moment of the star. We choose this initial amplitude such that the \alfven{} wave packet successfully breaks out from the magnetosphere at $r\sim 10r_{\rm in}$. We also ran a simulation with half the initial amplitude; however, the wave packet was not able to launch an ejecta in that case. We focus on the first case below.

\subsection{Nonlinear evolution of the \alfven{} wave}

The perturbation at the inner boundary launches a torsional \alfven{} wave (and a small amount of fast magnetosonic wave, see Appendix \ref{sec:fastwave}). For each half wavelength, the current structure consists of a core, aligned or anti-aligned with the background magnetic field, surrounded by a return current sheath of finite thickness. The field lines within the flux tube perturbed by the \alfven{} wave experience an alternation between clockwise and counterclockwise twisting.

The wave initially propagates along the dipole field lines and its relative amplitude grows with radius as $r^{3/2}$. The wave packet becomes significantly nonlinear at $r\sim10$. It is no longer confined to dipole field lines, but instead moves radially outward. 
The wave packet pushes the dipole field lines to open up, then the stretched field lines start to reconnect near the equator, allowing the twisted field lines in the wave packet to start detaching from the dipole magnetosphere.
The wave packet is therefore launched as an ejecta.
We take $r=10$ as the ejection radius, $R_{\rm ej}$.
Figure \ref{fig:Bslice} shows a few snapshots of the magnetic field and electric field on the $\phi=0$ plane, which cuts through the azimuthal center of the wave packet.
Figure \ref{fig:dUB_3d} shows the 3D structure of the perturbation electromagnetic energy density $\delta U$ and magnetic field lines within the flux bundle perturbed by the \alfven{} wave. $\delta U$ is defined as $\delta U=(\delta \mathbf{B}^2+\delta\mathbf{E}^2)/2$, where $\delta\mathbf{B}=\mathbf{B}-\mathbf{B}_0$, $\mathbf{B}_0$ is the initial background magnetic field, and $\delta\mathbf{E}=\mathbf{E}-\mathbf{E}_0=\mathbf{E}$.
From the $E_y$ plot in Figure \ref{fig:Bslice}, and the tenuous, spherical shell-like structure in Figure \ref{fig:dUB_3d}, it can be seen that a low frequency fast magnetosonic wave is generated at the leading edge of the ejecta; this is the consequence of the nonlinear conversion of \alfven{} mode to fast mode when the \alfven{} wave propagates along curved background magnetic field lines.

As the ejecta moves out, its thickness $\Delta r$ remains the same, but it expands laterally, roughly following spherical expansion from the star, so that the solid angle spanned by the ejecta remains more or less the same. The constant thickness can be understood from the conservation of magnetic energy and flux. Energy conservation requires $B^2r^2\Delta r={\rm const}$. Since the field in the ejecta is mostly transverse (in the $\theta$, $\phi$ directions), the flux conservation can be written as $Br\Delta r={\rm const}$. The two conditions then suggest $\Delta r={\rm const}$ and $B\propto r^{-1}$. We will confirm this scaling relation in the following subsection. In addition, within the ejecta, each half wavelength moves slightly sideways following its twisting direction. The ejecta looks like displaced, stacked pancakes, as shown in Figure \ref{fig:Bphi3d}, where we plot the 3D structure of the $B_{\phi}$ component at a particular time step $t=20$. 

After the ejection, at $t=25$ when the wave packet has reached $r\approx 2R_{\rm ej}$, we find that about half of the initial \alfven{} wave energy resides in the ejecta. For the rest of the energy, a significant fraction is used to push on the background field lines, stretching them out radially. We can roughly estimate how much work is done by the \alfven{} wave packet on the background magnetosphere as follows. When the ejecta has moved to a radius $r>R_{\rm ej}$, between $R_{\rm ej}$ and $r$, the initial dipole field is stretched into a monopole-like field. 
This takes energy per unit solid angle 
\begin{align}
    \Delta E&=E_{\rm monopole}-E_{\rm dipole}\nonumber\\
    &=\int_{R_{\rm ej}}^r\frac{1}{8\pi}\left(\frac{B_{\rm ej}R_{\rm ej}^2}{r^2}\right)^2r^2dr-\int_{R_{\rm ej}}^r\frac{1}{8\pi}\left(\frac{B_{\rm ej}R_{\rm ej}^3}{r^3}\right)^2r^2dr\nonumber\\
    &=\frac{B_{\rm ej}^2R_{\rm ej}^3}{8\pi}\left(1-\frac{R_{\rm ej}}{r}\right)-\frac{B_{\rm ej}^2R_{\rm ej}^3}{3\times8\pi}\left(1-\frac{R_{\rm ej}^3}{r^3}\right).
\end{align}
For our case, taking $r=2R_{\rm ej}$, and the total solid angle within which the field lines open up is $\Omega_{\rm total}\sim 2$, we find that the work done is 
\begin{equation}
    \Delta E\sim\frac{5}{24}\frac{B_{\rm ej}^2R_{\rm ej}^3}{8\pi}\Omega_{\rm total}
\end{equation}
This is a fixed amount regardless of the initial \alfven{} wave energy. It turns out to be about a quarter of the initial \alfven{} wave energy in our case.
This energy is stored in the stretched field lines, and some of it is dissipated in the formed current sheet due to magnetic reconnection. Besides this, there is also some energy that follows a portion of the \alfven{} wave to go back toward the southern pole of the star before the ejection happens, which starts to bounce back and forth in the inner magnetosphere and gradually gets dissipated. 

Figure \ref{fig:dUangular} shows the angular distribution of the electromagnetic energy in the ejecta. The 60\% containment region has a solid angle $\Omega\sim0.5$ steradian and the 80\% containment region has a solid angle $\Omega\sim1$ steradian. We note that at the inner boundary $r=r_{\rm in}$, the \alfven{} wave perturbation has an angular size $\Omega_0\sim0.12$. The wave first evolves linearly along the background dipole magnetic field, so the angular size grows with radius: $\Omega\sim\Omega_0r/r_{\rm in}$. Near the ejection radius $R_{\rm ej}\sim10r_{\rm in}$, we should have $\Omega\sim1.2$. After the ejection, the ejecta roughly follows spherical expansion from the star, so the angular distribution remains more or less the same. Our measured angular size is indeed consistent with this picture.

\begin{figure*}
    \centering
    \includegraphics[width=0.32\textwidth]{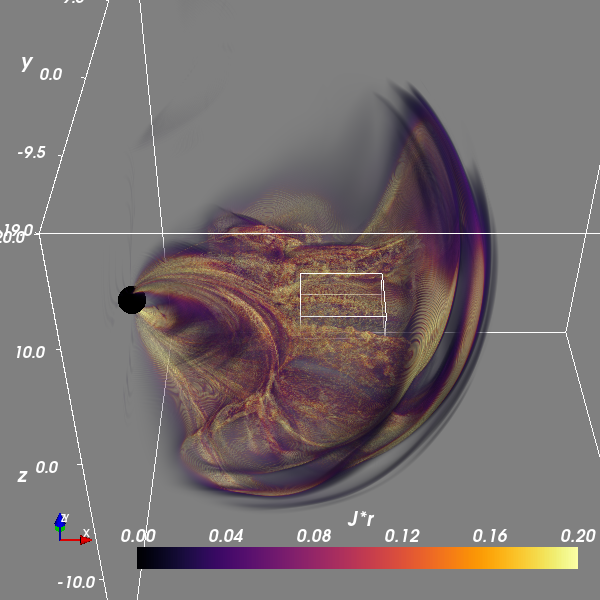}
    \includegraphics[width=0.32\textwidth]{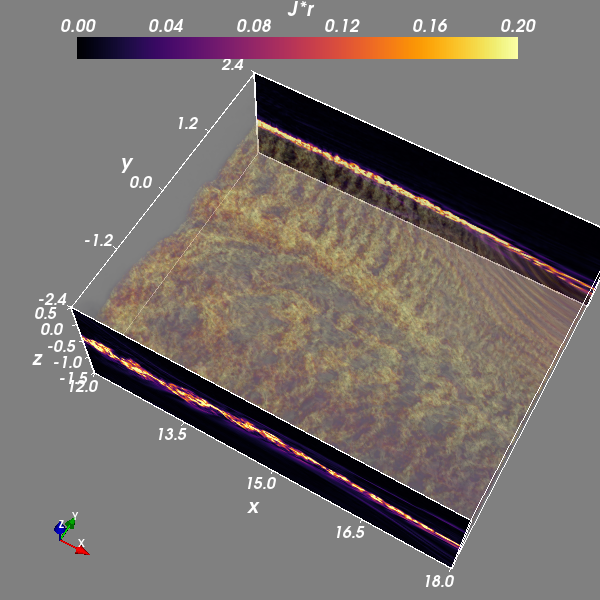}
    \includegraphics[width=0.32\textwidth]{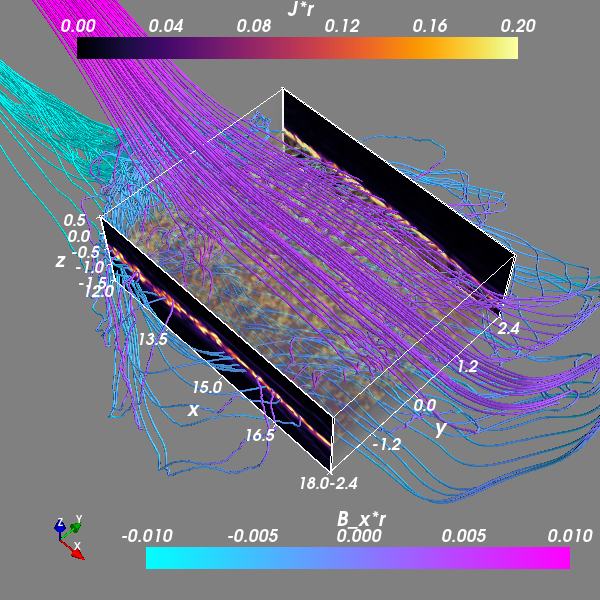}
    \caption{Left panel shows the 3D volume rendering of the magnitude of the current density in the global simulation, at the time step $t=30$. Middle panel shows a zoom-in view of the equatorial current sheet; the region corresponds to the small box in the left panel. Right panel shows the same region as the middle panel, where we also plot the field lines near the current sheet. The field lines are colored according to the value of $B_x r$.}
    \label{fig:J3d}
\end{figure*}

\begin{figure*}
    \centering
    \includegraphics[width=0.8\textwidth]{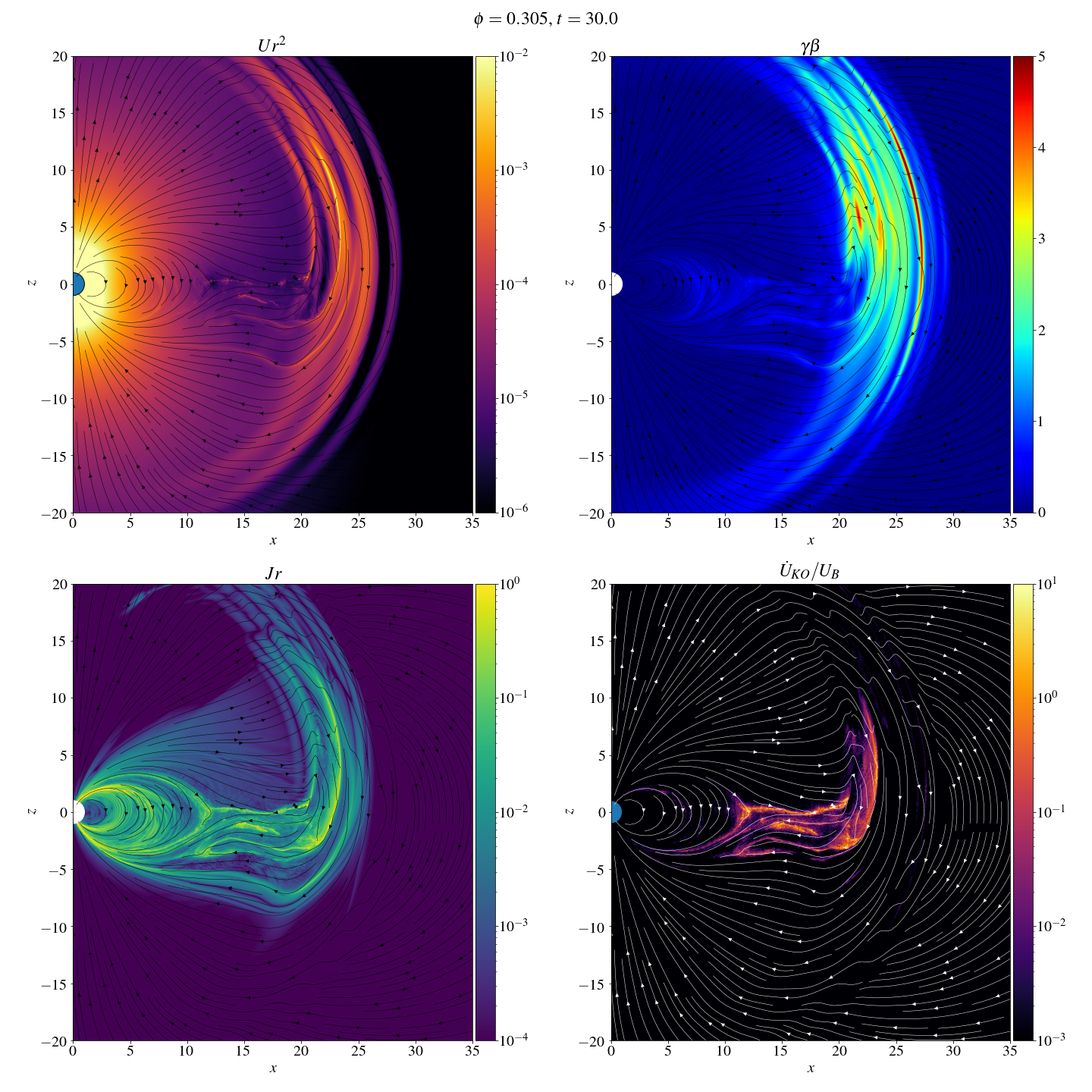}
    \caption{A slice on the $\phi=0.305$ plane at time $t=30$, showing the energy density of the electromagnetic field $U=(B^2+E^2)/2$, the fluid proper velocity $\gamma\beta$ calculated using the $\mathbf{E}\times\mathbf{B}$ drift (the proper velocity $\gamma\beta$ has been smoothed using a Gaussian kernel with a standard deviation of $24\Delta x$, where $\Delta x=1/128$ is the grid resolution), the current density, and the numerical dissipation rate $\dot{U}_{\rm KO}$ weighted by the local magnetic energy density.}
    \label{fig:UdUgm}
\end{figure*}

\begin{figure*}
    \centering
    \begin{minipage}{0.45\textwidth}
        \includegraphics[width=\textwidth]{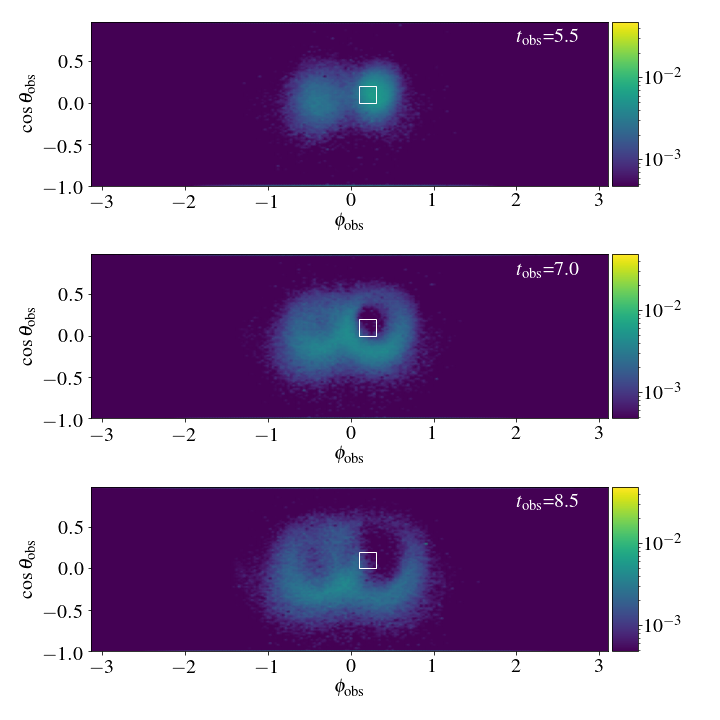}
    \end{minipage}
    \begin{minipage}{0.45\textwidth}
        \includegraphics[width=\textwidth]{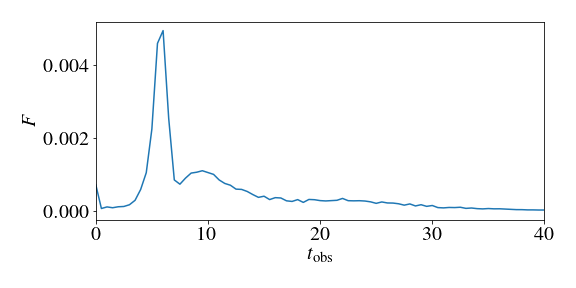}\\
    \includegraphics[width=\textwidth]{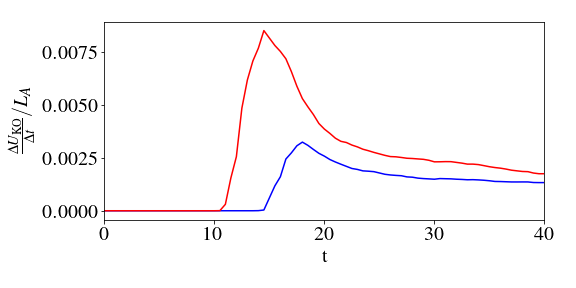}
    \end{minipage}
    \caption{Left: sky map of the X-ray emission, at three different observer times. We use $\dot{U}_{KO}/U_B$ as a proxy for the emissivity. We include emission from regions outside $r=5$. The flux shown is in arbitrary units. Right top panel: light curve of the X-ray emission within the small white box in the left panel. $F$ is the flux averaged over the small white box, in arbitrary units. Right bottom panel: total energy dissipation rate as a function of time. Red line shows the energy dissipation rate outside $r=5$, while the blue line shows that outside $r=10$. In both panels, times are in units of $r_{\rm in}/c=10^7\,{\rm cm}/c=0.33\,{\rm ms}$.}
    \label{fig:skymap_lightcurve}
\end{figure*}

\subsection{Scaling of quantities in the ejecta}

To better understand how the ejecta evolves, we measure the scaling of a few quantities in the ejecta. Firstly, we look at the peak electromagnetic field in the ejecta. A convenient measure to use is the peak perturbation energy density $\delta U$. We track the location $\mathbf{r}_m$ of the maximum $\delta Ur^2$ at each output time step; we make sure that the maximum is located in the main ejecta instead of the current sheet by choosing the maximum location on a data set smoothed using a Gaussian kernel with a standard deviation of $24\Delta x$, where $\Delta x=1/128$ is the grid resolution. Figure \ref{fig:Um_ud_scaling} shows the result. On the left panel, we show the distance $r_m$ as a function of time. This shows that the peak point is indeed consistently located on one peak/trough of the \alfven{} wave packet. In fact, the velocity of the pattern deviates slightly from a purely radial direction, and the speed is indistinguishable from $c$. The middle panel shows $\delta U_m\equiv\delta U(\mathbf{r}_m)$ as a function of $r_m$. We can see that $\delta U_m$ is consistent with decreasing as $r_m^{-2}$ besides some additional dissipation. This confirms that the magnetic field in the ejecta scales approximately as $r^{-1}$, instead of $r^{-2}$ as suggested by \citet{2021MNRAS.tmp.2876L}.

Now let us turn to the fluid velocity in the ejecta. 
We can define a velocity field $\mathbf{v}=\mathbf{E}\times\mathbf{B}/B^2$, then $\mathbf{E}=-\mathbf{v}\times\mathbf{B}$, and the force-free equations (\ref{eq:FF_dEdt})-(\ref{eq:FF_J}) can be cast into the following form \citep[e.g.,][]{1999astro.ph..2288G}
\begin{align}
    \frac{\partial\mathbf{B}}{\partial t}&=\nabla\times(\mathbf{v}\times\mathbf{B}),\\
    \frac{\partial}{\partial t}(B^2\mathbf{v})&=(\nabla\times\mathbf{B})\times\mathbf{B}+(\nabla\times\mathbf{E})\times\mathbf{E}+(\nabla\cdot\mathbf{E})\mathbf{E}.
\end{align}
This set of equations is very similar to the usual MHD equations, except that the inertia is provided by $B^2$. Here $\mathbf{v}$ is essentially the plasma drift velocity. In force-free electrodynamics, the fluid velocity itself is not defined and cannot be obtained directly from the fields, as there can be an arbitrary velocity component along $\mathbf{B}$. However, the drift velocity can be a good reference to gain insights into the plasma motion. In what follows, we look into the evolution of this drift velocity $\mathbf{v}$ in the ejecta. The corresponding Lorentz factor is $\gamma=1/\sqrt{1-v^2}$, and the proper velocity is $u=\gamma v$. 
Figure \ref{fig:Um_ud_scaling} right panel shows the evolution of $u$ and its components at the point of the maximum perturbation electromagnetic energy density. The velocity field is also smoothed using a Gaussian kernel with a standard deviation of $24\Delta x$. It can be seen that after the ejection, namely, after $t\sim15$, the drift proper velocity grows more or less linearly, and at large distances, this drift velocity is mostly radial. Another important point to note is that, although the fluid velocity at the peak/trough of the \alfven{} wave packet can be quite large, the \alfven{} wave packet is still a smooth wave structure even after the ejection: fluid enters from the front of the wave, then exits from behind. Shock formation is possible at large distances; this requires taking into account the inertia of the fluid and going beyond force-free approximation. Including the fluid inertia and pressure may also change the acceleration history of the ejecta. We leave this to future studies.

\subsection{Magnetic reconnection and dissipation}

When the ejection happens, magnetic field lines in the sheared flux bundle and ahead of it are pushed open by the \alfven{} wave packet. 
The left panel of Figure \ref{fig:J3d} shows the resulting current distribution in the magnetosphere. Behind each of the half-wavelength pancakes in the ejecta, there are current layers (which look more like current filaments) connecting the pancake with the closed zone. A main current sheet forms near the equatorial plane where opposite open magnetic fluxes from the northern and southern hemispheres meet. This is also seen in the third panel of Figure \ref{fig:Bslice}. Plasmoid-mediated reconnection happens in the equatorial current sheet (middle and right panel of Figure \ref{fig:J3d}), allowing the open field lines to reconnect and return to the closed initial state. 
From the right panel of Figure \ref{fig:J3d}, we can clearly see that it is primarily the poloidal field component that is reconnecting at the equatorial current sheet. 

Dissipation of the electromagnetic energy can happen at the current sheets. In our simulation, the dissipation is numerical and occurs through three channels: (1) Kreiss-Oliger dissipation that filters out the high frequency noise; (2) when $E>B$, $E$ is reduced to $B$; (3) when $\mathbf{E}\cdot\mathbf{B}\ne 0$, the component of $\mathbf{E}$ that is parallel to $\mathbf{B}$ is cut away. It turns out that most of the dissipation is accounted for by the first channel, the Kreiss-Oliger dissipation. In figure \ref{fig:UdUgm}, we show where this dissipation is triggered in a snapshot. It can be seen that the dissipation is concentrated along the current sheets. This is indeed consistent with our expectation that current sheets are natural sites for energy dissipation. These are likely sites for efficient X-ray emission.

We use the numerically dissipated energy as a proxy, to provide a picture of how the X-ray light curve behaves.
In particular, since the physical dissipation preferably happens at reconnection sites where the magnetic field changes direction significantly, these reconnection sites tend to have weaker magnetic field compared to other types of dissipation sites. So we use $\dot{U}_{KO}/U_B$ as a proxy for the emissivity, where $U_B=B^2/2$ is the local magnetic energy density. 
We assume that the emission is isotropic in the fluid rest frame; we also assume that the fluid moves with the drift velocity $\mathbf{v}=\mathbf{E}\times\mathbf{B}/B^2$, so the beaming of the received emission is affected by the fluid velocity. We calculate the sky map as a function of the observer angle and the observer time, taking into account the light travel time across the simulation box. This is done using a Monte-Carlo approach: we assign 1--2 particles per grid cell; for each particle, the emissivity is assigned to be $\dot{U}_{KO}/U_B$ in the lab frame, and the beaming direction is randomly drawn from an isotropic distribution in the fluid frame then boosted into the lab frame using the fluid drift velocity.
Figure \ref{fig:skymap_lightcurve} left panel shows a few snapshots of the sky map, at different observer times. It can be seen that the emission first beams around the equator, then expands and moves downward. It turns out that most of this beamed emission comes from the portion of the current sheet within the ejecta, namely, the vertical current sheet that shows up in the bottom panels of Figure \ref{fig:UdUgm}. This part moves relativistically with the ejecta; its Lorentz factor is already a few at $r\sim 2R_{\rm ej}$, as shown in the upper right panel of Figure \ref{fig:UdUgm}. This results in beamed X-ray emission. We see two peaks offset in $\phi$ angle in the sky map; this is because the half cycles in the \alfven{} wave with different twisting directions move slightly sideways with respect to each other (Figure \ref{fig:Bphi3d}), and the emission from the current sheet within each of the half cycles also beams differently.

In Figure \ref{fig:skymap_lightcurve} right panel, we show the light curve at a particular observer angle, corresponding to the small white box in Figure \ref{fig:skymap_lightcurve} left panel. As a comparison, we also show the total dissipated energy in the simulation box, as a function of the simulation time. Although the overall dissipation happens on a time scale $\sim R_{\rm ej}/c$, the observed light curve is much more peaked. This is also due to the relativistic effect: as the ejecta moves relativistically toward the observer, the arrival time of the emission is compressed by a factor $(1-\beta)$.   

\section{Discussion}\label{sec:discussion}

Firstly, considering the energetics, if we scale our simulation to realistic parameters of SGR 1935+2154, the stellar magnetic field is $B_0=4.4\times10^{14}$ G at the pole, and the ejection radius is $R_{\rm ej}=10^{8}$ cm, then the injected \alfven{} wave packet has an energy $\mathscr{E}_A=1.3\times10^{40}$ erg, and the initial relative amplitude of the \alfven{} wave is $\delta B/B\sim 10^{-3}$ at the maximum. As a comparison, the background magnetospheric energy at $R_{\rm ej}$ is roughly $\mathscr{E}_{\rm bg}\sim B^2R_{\rm ej}^3/(8\pi)\sim7.7\times10^{39}$ erg, so the \alfven{} wave can successfully break out from the magnetosphere. Another run we did with half the perturbation magnitude, thus $1/4$ of the energy in the \alfven{} wave packet, $\mathscr{E}_A\sim 3\times 10^{39}$ erg, did not successfully produce an ejecta. $\mathscr{E}_A$ must be well above $\mathscr{E}_{\rm bg}$ for the nonlinear wave packet to overcome confinement by the surrounding background field. This threshold is comparable to that found in our axisymmetric simulations \citep{2020ApJ...900L..21Y}, 
although it is somewhat lower, because the 3D \alfven{} wave packet needs to push open only a portion of the magnetosphere to break out.

A few features are robust across 2D and 3D simulations.
Although the initial \alfven{} wave perturbations are different in the 2D and 3D models, the ejecta structure on a poloidal plane looks remarkably similar.
The ejecta is mainly composed of the current carrying, twisted field of the \alfven{} wave packet, plus a fast wave in front of it, generated as the initial \alfven{} wave propagates along curved background field lines. After the ejection, in both 2D and 3D, the ejecta retains its radial thickness and solid angle, expands balistically from the star, and becomes a pancake-like structure at large distances. As the ejecta pushes open the magnetospheric field lines, the main current sheet is formed near the equatorial plane behind the ejecta.

However, the 3D nature of the initial \alfven{} wave perturbation and its subsequent evolution does produce a few new features. First, the angular distribution of the ejecta energy is not axisymmetric in the 3D model.
The angular size is ultimately determined by the disturbed region on the stellar surface that launches the \alfven{} wave. The wave initially evolves linearly along the background dipole field lines, its angular size growing proportional to $r$; after the ejection, the angular size becomes frozen. A compact wave launching region can thus produce an ejecta compact in angular size. Secondly, the current distribution shows a more complex structure in the case of the localized 3D wave launching.
Besides the equatorial current sheet, there are quite a few current filaments, especially near the lateral boundary of the perturbed magnetosphere. 

We also find that at the majority of the reconnecting current sheets, it is the poloidal component of the magnetic field that reconnects, not the transverse field in the \alfven{} wave. Although most of the magnetic energy in the \alfven{} wave initially resides in the transverse component, the wave packet gives part of its energy to the poloidal component by deforming the background magnetic field. The deformed poloidal magnetic field then reconnect and dissipates the energy. 

Our simulations are carried out in the FFE limit, neglecting the plasma inertia and pressure effects. As a result, there are only two characteristic wave modes, the \alfven{} mode and the fast mode, both having a group speed of $c$. Therefore, shocks cannot form in the FFE framework. To understand physically how the ejecta accelerates with radius, and how the shock forms as the ejecta runs into the magnetar wind, one would need to go beyond force-free approximation. This will be investigated using relativistic magnetohydrodynamics simulations in the future.

In addition, our force-free simulations cannot capture the microphysics of the dissipation processes happening at the current sheet and other locations. In reality, the reconnection physics at the current sheet is governed by 
kinetic plasma processes. Furthermore,
due to the relatively strong magnetic field and small length scales, the resulting radiation field is compact and the plasma strongly interacts with the radiation. Photons initially emitted through synchrotron radiation can experience additional inverse Compton scatterings and photon-photon pair production; photons can also be regenerated through pair annihilation. These 
processes will influence the plasma dynamics and
shape the emergent radiation spectrum
\citep{2021ApJ...921...92B}. 
Kinetic plasma simulations including all the relevant radiative processes are needed 
for a complete description of the reconnection process.

\section{Conclusion}\label{sec:conclusion}
We have carried out fully 3D force-free electrodynamics simulations of a localized \alfven{} wave packet launched by a magnetar quake into the magnetosphere. We find that if the \alfven{} wave packet propagates to a radius $R$ and has a total energy greater than the magnetospheric energy $B^2R^3/(8\pi)$, then the wave can become quite nonlinear and get ejected from the magnetosphere.
The ejecta can carry a large portion of the initial \alfven{} wave energy. 
The ejecta preserves its radial thickness during its
expansion from the star, so it becomes a pancake-like structure. 
Its angular size $\Omega$ is determined by the initial \alfven{} wave perturbation at the stellar surface: $\Omega\sim\Omega_0R_{\rm ej}/r_*$, where $\Omega_0$ is the perturbation solid angle at the stellar surface, and $R_{\rm ej}$ is the ejection radius. 
The ejecta pushes open the magnetospheric field lines, creating current sheets behind it that connect back to the closed zone. Magnetic reconnection can happen at these current sheets; this will lead to plasma energization and X-ray emission. The energy source of this dissipation is the magnetic energy contained in the stretched poloidal field lines. Some of the current sheets move relativistically with the ejecta; they can produce beamed X-ray emission, and may be responsible for the sharp spikes coincident with the radio bursts from SGR 1935+2154.

\bigskip
\begin{center}
    ACKNOWLEDGMENTS
\end{center}

We thank Bart Ripperda and Jens Mahlmann for insightful discussions, and the anonymous referee for helpful comments. Y. Y. is supported by a Flatiron Research Fellowship at the Flatiron Institute, Simons Foundation. A.M.B. is supported by grants from NSF AST-1816484 and AST-2009453, NASA 21-ATP21-0056, and Simons Foundation \#446228. A. C. is supported by NSF grants AST-1806084, AST-1903335, and acknowledges support from the Fermi Guest Investigation grant 80NSSC21K2027. Y. L. is supported by NSF grant AST-2009453. E. R. M. gratefully acknowledges support from postdoctoral fellowships at the Princeton Center for Theoretical Science, the Princeton Gravity Initiative, and the Institute for Advanced Study. A. P. is supported by NSF grant AST-1909458. This research is part of the Frontera \citep{Stanzione2020} computing project at the Texas Advanced Computing Center (LRAC-AST21006). Frontera is made possible by National Science Foundation award OAC-1818253. This research also used resources of the Oak Ridge Leadership Computing Facility, which is a DOE Office of Science User Facility supported under Contract DE-AC05-00OR22725. Research at the Flatiron Institute is supported by the Simons Foundation.

%



\software{{\it Coffee}, \url{https://github.com/fizban007/CoffeeGPU}, \citet{2020ApJ...893L..38C}
          }



\appendix
\section{Convergence of the force-free code {\it Coffee}}\label{sec:convergence}

{\it Coffee} \citep{2020ApJ...893L..38C} uses an algorithm similar to
\citet{2015PhRvL.115i5002E, 2016ApJ...817...89Z}: we use fourth-order
central finite difference stencils on a uniform Cartesian grid and a
five-stage fourth-order low storage Runge-Kutta scheme for time
evolution \citep{carpenter_fourth-order_1994}. We use hyperbolic
divergence cleaning \citep{2002JCoPh.175..645D} to damp any violations
of $\nabla\cdot\mathbf{B}=0$.
To enforce the force-free condition, we explicitly remove any
$\mathbf{E}_{\parallel}$ by setting
$\mathbf{E}\to\mathbf{E}-(\mathbf{E}\cdot\mathbf{B})\mathbf{B}/B^2$ at
every time step, and whenever $E>B$ happens, we reset $\mathbf{E}\to\mathbf{E}(B/E)$. We apply standard sixth order Kreiss-Oliger numerical
dissipation to all hyperbolic variables to suppress high frequency noise
from truncation error \citep{kreiss_methods_1973}:
\begin{equation}
    \partial_t U^{(new)}=\partial_t U+\epsilon_{\rm KO}\frac{1}{64}\left(\frac{\partial^6}{\partial x^6}+\frac{\partial^6}{\partial y^6}+\frac{\partial^6}{\partial z^6}\right)U,
\end{equation}
where $U$ represents any hyperbolic variables, $\epsilon_{\rm KO}<1$ is a constant parameter, and we use a second order stencil for the sixth order derivative.
The code is parallelized and optimized to run on GPUs as well as CPUs with excellent scaling.

We carried out convergence tests for the code {\it Coffee}, following procedures discussed by \citet{2021A&A...647A..58M}. We show the results from two most important tests below.

\begin{figure*}
    \centering
    \includegraphics[width=0.45\textwidth]{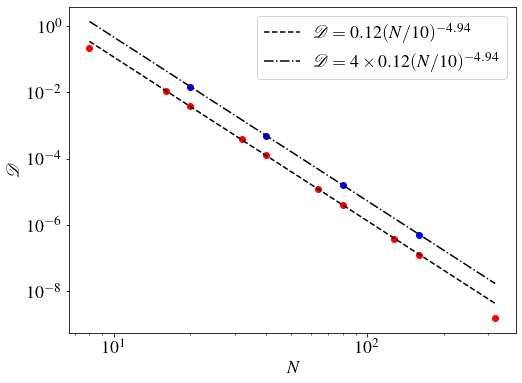}
    \includegraphics[width=0.45\textwidth]{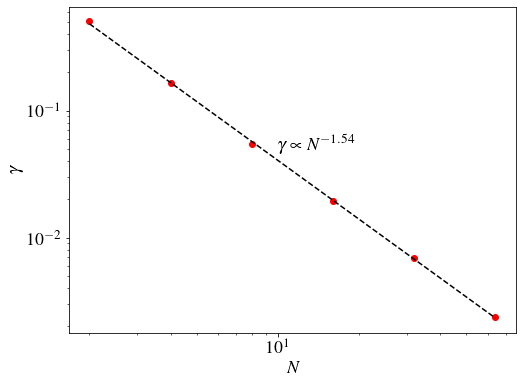}
    \caption{Left: convergence of the \alfven{} wave test. The data points are the measured damping rate $\mathscr{D}$ (in units of $c/L$) of the \alfven{} wave magnetic field as a function of the number of grid points $N$ on each side of the simulation box. The red dots correspond to runs with $\epsilon_{\rm KO}=0.1$, and the blue dots correspond to $\epsilon_{\rm KO}=0.4$. The dashed line is fitted to the red points. Right: convergence of the tearing mode test. The data points are the measured tearing mode growth rate $\gamma$ for a series of runs with different resolution. $N$ is the number of points resolving the current sheet thickness scale $a$.}
    \label{fig:convergence}
\end{figure*}

\subsection{Planar \alfven{} wave test}\label{subsec:convergence_alfven}
In this test, we set up a 3D Cartesian periodic box with size $L\times L\times L$, and a uniform background magnetic field $\mathbf{B}_0$ along the x direction. We initialize a planar \alfven{} wave, with wave vector $\mathbf{k}=(2,0,1)2\pi/L$, and relative amplitude $\xi=\delta B/B_0=0.1$. We let the wave evolve for a long time. Due to numerical diffusion, The wave magnetic field will slowly decay with time according to \begin{equation}
    \delta\mathbf{B}=\delta\mathbf{B}_i e^{-\mathscr{D}t},
\end{equation}
where $\delta\mathbf{B}_i$ is the magnetic field of the ideal wave solution in the absence of any numerical diffusion, and $\mathscr{D}$ is the damping rate. In force-free codes, $\mathscr{D}$ is related to the numerical resistivity through
\begin{equation}
    \mathscr{D}=\frac{k^2}{2}\eta,
\end{equation}
where $k$ is the wave vector, $\eta$ is the numerical resistivity \citep[][and references therein]{2021A&A...647A..58M}. The numerical resistivity can be written in the form
\begin{equation}
    \eta=\mathcal{R}\mathcal{V}\mathcal{L}\left(\frac{\Delta x}{\mathcal{L}}\right)^r,
\end{equation}
where $\mathcal{R}$ is a resolution independent numerical coefficient, $\mathcal{L}$ and $\mathcal{V}$ are the characteristic length and speed of the problem, $\Delta x$ is the grid spacing, and $r$ is the measured order of convergence. 

In this \alfven{} wave test, we only change the grid resolution, namely $\Delta x$, to measure the damping rate $\mathscr{D}$ and the order of convergence $r$. The damping rate is more conveniently measured using the total wave energy $\delta \mathcal{E}$:
\begin{equation}
    \delta \mathcal{E}=\delta \mathcal{E}_ie^{-2\mathscr{D}t}.
\end{equation}
We use a simulation grid with a number of $N^3$ points, where $N$ is the number of cells on each side of the box, which ranges from 16 to 320 in the series of simulations. Figure \ref{fig:convergence} left panel shows the measured $\mathscr{D}$ from these runs. We can see that the order of convergence $r$ is around 5 for our scheme. It turns out that the Kreiss-Oliger dissipation is one of the most important source for the numerical resistivity; it seems to determine the convergence order. The wave damping rate $\mathscr{D}$ also directly depends on the prefactor $\epsilon_{\rm KO}$ of the Kreiss-Oliger dissipation term. Figure \ref{fig:convergence} left panel shows a direct proportionality between $\mathscr{D}$ and $\epsilon_{\rm KO}$. We use $\epsilon_{\rm KO}=0.1$ for the global simulations presented in the paper. We can also see that the wave damping rate is less than $10^{-2}c/L$ when there are more than 16 grid points per side of the box, or more than 8 points per wavelength.

\subsection{Tearing mode test}\label{subsec:convergence_tearing}
In this test, we set up a force-free current sheet similar to \citet{2021A&A...647A..58M}. Our simulation box has a length of L=2 along $x$ and $y$ directions, and a length of 3L=6 along z direction. The background magnetic field has the following form
\begin{align}
    B_{0x}&=B_0\tanh(z/a),\nonumber\\
    B_{0y}&=B_0\,\mathrm{sech}(z/a),
\end{align}
and we set $a=0.1$. The field is initially perturbed by \begin{align}
    B_{1x}&=\epsilon (ak)^{-1}B_0\sin(kx)\tanh(z/a)\,\mathrm{sech}(z/a),\nonumber\\
    B_{1z}&=\epsilon B_0\cos(kx)\,\mathrm{sech}(z/a),
\end{align}
where $k=2\pi/L$ is the perturbation wavenumber. We set the perturbation amplitude to be $\epsilon=10^{-4}$. The boundary condition is periodic in $x$ and $y$ directions, and has zero derivative in the $z$ direction. The growth rate of the tearing mode can be traced using the $B_z$ component, which grows exponentially with time: $B_z=B_z(t=0)e^{\gamma t}$. Figure \ref{fig:convergence} right panel shows the measured tearing mode growth rate for a series of runs with different resolutions. We find that roughly $\gamma\propto N^{-1.54}$, where $N$ is the number of grid points within the current sheet thickness scale $a$.

In resistive MHD description of the tearing mode, the growth rate of a single $k$ tearing mode is given by \citep[e.g.,][]{2017ApJS..230...18R,2021A&A...647A..58M}
\begin{equation}
    \gamma=1.06^{-4/5}\eta^{3/5}v_A^{2/5}a^{-8/5}(ak)^{2/5}\left(\frac{1}{ak}-ak\right)^{4/5},
\end{equation}
where $\eta$ is the resistivity, and $v_A$ is the \alfven{} speed.
On the other hand, the growth rate of the fastest-growing mode is \citep{1963PhFl....6..459F}
\begin{equation}
    \gamma_{\rm max}\approx 0.6a^{-3/2}v_A^{1/2}\eta^{1/2}.
\end{equation}
Suppose the growth rate we measured is the maximum growth rate, then we would obtain the relation between the resistivity and the grid resolution as $\eta\propto N^{-3.08}$. This order of convergence seems to be different from what we found in \S\ref{subsec:convergence_alfven}. This is because in the tearing mode experiment, there can be locations where the force-free condition is violated, therefore the enforcement of force-free condition is activated and the components of the electric field that violate $E<B$ or $\mathbf{E}\cdot\mathbf{B}=0$ are cut away. This will affect the actual numerical resistivity. We expect the convergence order to follow the order of the time integration in this case. The conclusion is similar to \citet{2021A&A...647A..58M}.

\section{Fast waves launched from the inner boundary}\label{sec:fastwave}

\begin{figure}
    \centering
    \includegraphics[width=0.6\columnwidth]{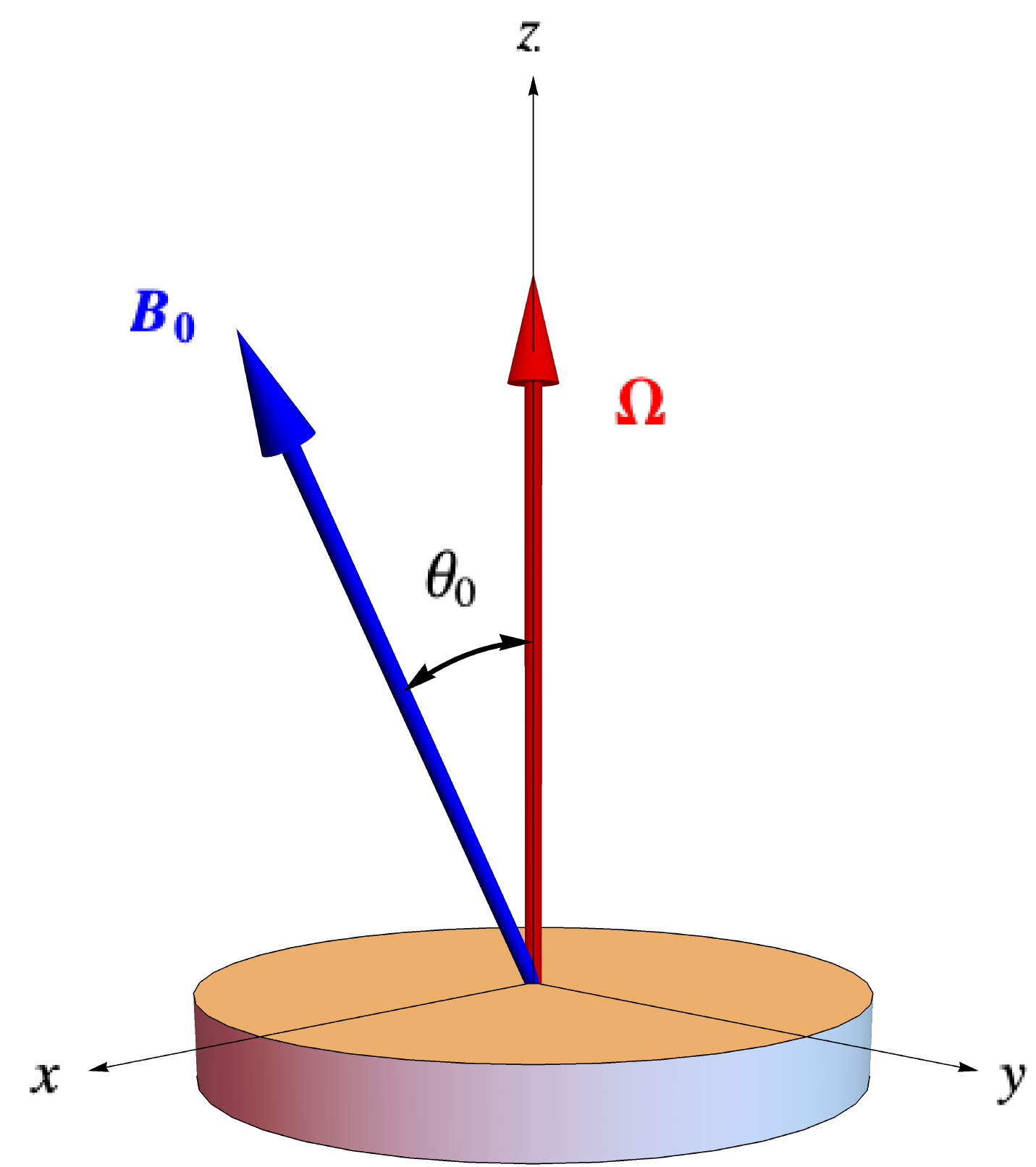}
    \caption{A simplified setup to understand the wave modes launched from the perturbation on the inner boundary of the global simulation.}
    \label{fig:disk}
\end{figure}

To understand the wave modes launched from the inner boundary, let us consider the following simplified problem. Consider an infinitely large conductor covering the space $z<0$, while the region $z>0$ is filled with a force-free plasma. There is a uniform magnetic field making an angle $\theta_0$ with respect to the normal of the conductor. Without loss of generality we assume that the magnetic field lies in the xz plane, $\mathbf{B}_0=B_0(\sin\theta_0\,\mathbf{\hat{x}}+\cos\theta_0\,\mathbf{\hat{z}})$. A circular region on the surface of the conductor is twisted with a radially dependent angular velocity $\pmb{\Omega}=\Omega(R,t)\mathbf{\hat{z}}$, as shown in Figure \ref{fig:disk}. On the surface of the rotating region, a point with cylindrical coordinates $(R,\phi,z=0)$ has the following velocity
\begin{equation}
    \mathbf{v}=v_{\phi}\pmb{\hat{\phi}}=-v_{\phi}\sin\phi\,\mathbf{\hat{x}}+v_{\phi}\cos\phi\,\mathbf{\hat{y}},
\end{equation}
where $v_{\phi}=\Omega(R)R$. The rotation induced electric field at this point is then
\begin{align}\label{eq:disk_E1}
    \mathbf{E}&=-\mathbf{v}\times\mathbf{B}_0\nonumber\\
    &=B_0v_{\phi}(-\cos\theta_0\mathbf{\hat{R}}+\sin\theta_0\cos\phi\,\mathbf{\hat{z}})\nonumber\\
    &\equiv \mathbf{E}_R+\mathbf{E}_z.
\end{align}
The magnitude of $\mathbf{E}$ is
\begin{equation}
    E=B_0|v_{\phi}|\sqrt{\cos^2\theta_0+\sin^2\theta_0\cos^2\phi},
\end{equation}
and for small $\theta_0$, $|E_z/E_R|\lesssim\tan\theta_0\ll1$.
Since the conductor is surrounded by a perfectly conducting plasma, immediately outside the conductor, the electric field should be continuous. To determine the nature of the modes, we carry out a local expansion of the electric field immediately outside the conductor around the point $(R,\phi,z=0^{+})$ into force-free normal modes. Since \alfven{} modes have $\mathbf{E}$ lying in the $\mathbf{k}$--$\mathbf{B}_0$ plane, while the fast modes have $\mathbf{E}$ perpendicular to the $\mathbf{k}$--$\mathbf{B}_0$ plane, we can find the component of the fast mode by projecting the electric field on to the normal of the $\mathbf{k}$--$\mathbf{B}_0$ plane.

We look at the $\mathbf{E}_R$ component first. It does not have $\phi$ dependence, therefore the wave vector only has $\mathbf{\hat{R}}$ and $\mathbf{\hat{z}}$ components: $\mathbf{k}=k_R\mathbf{\hat{R}}+k_z\mathbf{\hat{z}}$. The unit vector along the normal of the $\mathbf{k}$--$\mathbf{B}_0$ plane can be written as
\begin{align}\label{eq:k_cross_B_unit}
    \mathbf{n}&=\frac{\mathbf{k}\times\mathbf{b}_0}{\left|\mathbf{k}\times\mathbf{b}_0\right|}
\end{align}
where $\mathbf{b}_0=\mathbf{B}_0/B_0$ is a unit vector along the background magnetic field,
\begin{align}
    \mathbf{k}\times\mathbf{b}_0&=k_R\cos\theta_0\sin\phi\,\mathbf{\hat{x}}\nonumber\\
    &+(-k_R\cos\theta_0\cos\phi+k_z\sin\theta_0)\,\mathbf{\hat{y}}\nonumber\\
    &+k_R\sin\theta_0\sin\phi\,\mathbf{\hat{z}},
\end{align}
\begin{align}
    \left|\mathbf{k}\times\mathbf{b}_0\right|&=\left(k_R^2\cos^2\theta_0-2k_Rk_z\sin\theta_0\cos\theta_0\cos\phi\right.\nonumber\\
    &\left.+\sin^2\theta_0(k_R^2\sin\phi^2+k_z^2)\right)^{1/2}.
\end{align}
The magnitude of the fast mode electric field is then
\begin{align}
    |\mathbf{E}_{f1}|=|\mathbf{E}_R\cdot\mathbf{n}|=\frac{B_0v_{\phi}k_z\sin\theta_0\cos\theta_0\sin\phi}{\left|\mathbf{k}\times\mathbf{b}_0\right|}.
\end{align}
Typically in our boundary condition, $k_R\gg k_z$, and $\theta_0\lesssim0.3$, so we can see that $|\mathbf{E}_{f1}|/|\mathbf{E}_R|\sim (k_z/k_R)\sin\theta_0\ll1$.

Now let us look at the $\mathbf{E}_z$ component in Equation (\ref{eq:disk_E1}). For this component, the wave vector does have $\phi$ dependence: $\mathbf{k}=k_R\mathbf{\hat{R}}+k_{\phi}\pmb{\hat{\phi}}+k_z\mathbf{\hat{z}}$.
The unit vector along the normal of the $\mathbf{k}$--$\mathbf{B}_0$ plane is still given by Equation (\ref{eq:k_cross_B_unit}), but with
\begin{align}
    \mathbf{k}\times\mathbf{b}_0&=\cos\theta_0(k_{\phi}\cos\phi+k_R\sin\phi)\mathbf{\hat{x}}\nonumber\\
    &+(k_z\sin\theta_0+\cos\theta_0(-k_R\cos\phi+k_{\phi}\sin\phi))\mathbf{\hat{y}}\nonumber\\
    &-\sin\theta_0(k_{\phi}\cos\phi+k_R\sin\phi)\mathbf{\hat{z}},
\end{align}
and its norm is
\begin{align}
    \left|\mathbf{k}\times\mathbf{b}_0\right|&=\left[\sin^2\theta_0\left(k_z^2+(k_{\phi}\cos\phi+k_R\sin\phi)^2\right) \right.\nonumber\\
    &+k_z\sin(2\theta_0)(-k_R\cos\phi+k_{\phi}\sin\phi))\nonumber\\
    &\left.+(k_R^2+k_{\phi}^2)\cos^2\theta_0\right]^{1/2}.
\end{align}
The magnitude of the fast mode electric field is
\begin{equation}
    |\mathbf{E}_{f2}|=|\mathbf{E}_z\cdot\mathbf{n}|=\frac{B_0v_\phi\sin^2\theta_0\cos\phi(k_{\phi}\cos\phi+k_R\sin\phi)}{\left|\mathbf{k}\times\mathbf{b}_0\right|}.
\end{equation}
As a result, $|\mathbf{E}_{f2}|/|\mathbf{E}_R|\sim\sin^2\theta_0$.

Putting together the above results, we can see that if $\theta_0=0$, namely, $\mathbf{B}_0$ is perfectly perpendicular to the conductor surface, the launched wave mode is purely \alfven{ic}. For small angle $\theta_0$, the fast mode electric field is a factor of $\max(\sin\theta_0 k_z/k_R,\sin^2\theta_0)$ compared to the total electric field. In our boundary condition for the global simulation, typically $k_z/k_R\ll0.1$, and $\theta_0\lesssim0.3$, therefore the fast mode electric field amplitude is at most 0.1 of the total electric field, and its energy is at most 1\% of the total perturbation. Furthermore, fast modes, unlike \alfven{} waves, are not collimated by the field lines and therefore propagate more or less isotropically out and decrease more quickly than \alfven{} waves; and in the case of our boundary perturbations, the two sides of the rotating region will create fast wave contributions that will be negatively interfering after the wave propagates far enough. Therefore, the effect of the fast waves launched from the boundary is negligible in our simulations.


\bibliography{ref}{}
\bibliographystyle{aasjournal}



\end{document}